# Surface activation by electron scavenger metal nanorod adsorption on TiH$_2$, TiC, TiN, and Ti$_2$O$_3$

Yoyo Hinuma[1,2*], Shinya Mine[3], Takashi Toyao[3,4], Zen Maeno[3], and Ken-ichi Shimizu[3,4]

[1] Institute of Innovative Research, Tokyo Institute of Technology, 4259 Nagatsuta-cho, Midori-ku, Yokohama, 226-8502, Japan
[2] Department of Energy and Environment, National Institute of Advanced Industrial Science and Technology (AIST), 1-8-31, Midorigaoka, Ikeda 563-8577, Japan
[3] Institute for Catalysis, Hokkaido University, N-21, W-10, 1-5, Sapporo 001-0021, Japan
[4] Elements Strategy Initiative for Catalysts and Batteries, Kyoto University, Katsura, Kyoto 615-8520, Japan

* y.hinuma@aist.go.jp

**Abstract**
Metal/oxide support perimeter sites are known to provide unique properties because the nearby metal changes the local environment on the support surface. In particular, the electron scavenger effect reduces the energy necessary for surface anion desorption, thereby contributes to activation of the (reverse) Mars-van Krevelen mechanism. This study investigated the possibility of such activation in hydrides, carbides, nitrides, and sulfides. The work functions (WFs) of known hydrides, carbides, nitrides, oxides, and sulfides with group 3, 4, or 5 cations (Sc, Y, La, Ti, Zr, Hf, V, Nb, and Ta) were calculated. The WFs of most hydrides, carbides, and nitrides are smaller than the WF of Ag, implying that the electron scavenger effect may occur when late transition metal nanoparticles are adsorbed on the surface. The WF of oxides and sulfides decrease when reduced. The surface anion vacancy formation energy correlates well with the bulk formation energy in carbides and nitrides, while almost no correlation is found in hydrides because of the small range of surface hydrogen vacancy formation energy values. The electron scavenger effect is explicitly observed in nanorods adsorbed on TiH$_2$ and Ti$_2$O$_3$; the surface vacancy formation energy decreases at anion sites near the nanorod, and charge transfer to the nanorod happens when an anion is removed at such sites. Activation of hydrides, carbides, and nitrides by nanorod adsorption and screening support materials through WF calculation are expected to open up a new category of supported catalysts.

## 1. Introduction

Metals and metal oxides play a central role in the field of heterogeneous catalysis.[1, 2] However, the application of alternative materials as catalysts is always important in exploring new processes and improving existing processes.[3] Among such materials are metal carbides, nitrides, sulfides and hydrides.[4-6] While these compounds are not new, there is growing interests in their catalytic properties and potential application that include thermal, electro- and photo-catalysis.[7-10] For instance, carbides and nitrides are known to play roles in a number of valuable catalytic processes such as Fischer–Tropsch synthesis,[11, 12] $NH_3$ synthesis,[13] hydroprocessing[14], and water splitting.[15] Sulfides have also been extensively used in the petroleum industry.[16, 17] Although hydrides have been less explored in heterogeneous catalysis, surface hydrides and hydride-containing mixed-anion compounds are receiving much attention recently because of potential applications to various catalytic processes, including $NH_3$ synthesis and $CO_2$ hydrogenation, where surface H species play important roles.[18-21]

One of the main interests of catalysis over the aforementioned materials is the reaction mechanism. In particular, the Mars–van Krevelen mechanism, where surface anion vacancies serve as active centers, is often considered.[22] Recently, Zeinalipour-Yazdi et al. investigated the mechanism of $NH_3$ synthesis over $Co_3Mo_3N$ using density functional theory (DFT) calculations and found that the $NH_3$ synthesis reaction can proceed via a N-based Mars–van Krevelen mechanism.[23] In addition, several studies reported that Fischer–Tropsch synthesis over Fe carbides proceed via the Mars–van Krevelen mechanism involving liberation of carbon from the carbide surface and dissociative adsorption of CO to fill the carbon vacancies to recover the carbide surface.[24-26] Although such works greatly contributed to the body of knowledge concerning catalysis of non-oxide-based materials, our present understanding of their catalytic roles and surface properties remains insufficient.

The interaction between the metal nanoparticle and its support has extensively been discussed in heterogeneous catalysis.[27-29] The supported metal nanoparticles and the supports affect each other, and they create unique surface properties and electronic structures.[30, 31] Sites located in the immediate vicinity of the metal/oxide interfaces, called perimeter sites, are often considered to be the catalytically active sites.[32-34] Although significant effort has been devoted to metal/oxide systems,[35, 36] the interaction of metal nanoparticles and non-oxide supports have been much less explored. Ye et al. recently showed that a Ni-loaded La nitride catalyst promotes $NH_3$ synthesis in which N vacancy sites of La nitride play essential roles.[37] Deposition of Ni facilitates the formation of the N vacancies (lower the N vacancy formation energy) by acting as electron scavengers, leading to the comparable $NH_3$ synthesis performance to that of Ru-based catalysts.[37] Despite significant potentials for future catalysis research, such metal/non-oxide support systems have not extensively been investigated due to its high complexity, and as a result, remain a key area of research in heterogeneous catalysis.[38, 39] Computational approaches to address this issue in a systematic fashion and to investigate underlying physics/chemistry are necessary as it is often difficult to access via experiments owing to the complex nature and catalytic properties of materials in principle should be determined by their surface electronic structures.

In this study, surface properties, namely the surface energy ($E_{surf}$) and the work function (WF), were calculated for binary compounds where the cation is a group 3, 4, or 5 element cation, namely Sc, Y, La, Ti, Zr, Hf, V, Nb, or Ta, and the anion is one of H, C, N, O, or S. Tables 1-5 show the compounds and phases investigated in this paper. Information on the compounds and phases were mostly obtained from binary phase diagrams in Ref. [40], and structural information, namely lattice parameters and atom positions, were primarily obtained from the Materials Project [41]. There were some cases where the crystal structure data in Ref. [40] and phases reported in the Materials Project were inconsistent, and there were also cases where phases from other sources were used instead. Handling of these situations are outlined in the notes of Tables 1-5. A total of nine hydrides, 13 carbides, 12 nitrides, 18 oxides, and 19 sulfides were considered. Trends in surface anion vacancy formation energies were investigated for fluorite or distorted fluorite structure hydrides as well as carbides and nitrides with the rocksalt structure. The effect of activation by adsorbing different metal nanorods, namely Re (group 7), Ru (group 8), Rh (group 9), Pd, Pt (group 10), Ag, Au (group 11), Zn (group 12), and Al (group 13), on activation of $TiH_2$, TiC, TiN, and $Ti_2O_3$ supports was also studied. Activation of anion sites, or reduction of surface anion vacancy formation energy ($E_{vac}$) as a neutral atom, was found when the nanorod WF was larger than the support. Activation was typically accompanied by charge transfer to the nanorod upon anion removal in $TiH_2$, TiN, and $Ti_2O_3$, therefore the activation is a result of the electron scavenger effect that was previously reported in oxides.[42, 43] For the record, the idea of charge transfer between an adsorbate and metal was discussed by Zheng et al., who studied interactions between C1 fragments and metal surfaces, in as early as 1988, which is decades earlier than the proposal of the electron scavenger concept.[44]

## 2. Methodology

First-principles calculations were conducted using the projector augmented-wave method[45] and the PBEsol[46] approximation to the exchange-correlation interactions as implemented in the VASP code.[47, 48] The PBEsol approximation reproduces experimental lattice parameters of elementary substances as well as $d^0$ and $d^{10}$ oxides well with few outliers. PBEsol can actually predict lattice parameters of low dimensional structures comparably well with PBED3[49] even though, unlike the latter, the former do not explicitly treat van der Waals interactions[50].

Slab-and-vacuum models, where slabs that infinitely extend in two directions (in-plane directions) are alternated with a vacuum region in the other direction under three-dimensional periodic boundary conditions, were obtained by considering low-index surfaces that are type 1 or 2 in Tasker's definition[51] or nonpolar type A or B in the definition by Hinuma *et al.*[52] Stoichiometric slabs, where the two surfaces are identical, can be obtained by simply cleaving bulk for nonpolar type A or B surfaces, and an automated procedure to generate such slabs exists.[52] Defect formation (anion desorption as well as metal nanorod adsorption) was performed on both sides of a slab such that the slab was always nonpolar. Internal coordinates and lattice parameters were relaxed in bulk calculations and, unless stated otherwise, all internal coordinates were allowed to relax while lattice parameters were fixed in slab calculations. Slabs were separated by at least ~12 Å in the direction perpendicular to the surface. In case of nanorod-adsorbed slabs, the positions of top and bottom of the nanorods were considered as the surface for this purpose.

The representative surface(s) of compounds in Tables 1-5 are shown in Figs. 1-2. The representative surface has the lowest $E_{surf}$ among investigated orientations and terminations, but two representative surfaces are considered when there are two surfaces with $E_{surf}$ within few meV/Å$^2$ of each other. Here, $E_{surf}$ is defined as

$$E_{surf} = (E_{slab} - E_{bulk})/2A$$

where $E_{slab}$ and $E_{bulk}$ are the energy of the slab without defects and the energy of the constituents of the slab when in a perfect bulk, respectively. $A$ is the in-plane area of the slab (the coefficient of 2 accounts for both sides of the slab). $E_{bulk}$ is obtained from a bulk calculation.

The anion vacancy formation energy $E_{vac}$ is defined as

$$E_{vac} = \left(E_{removed} - E_{original} + \mu_i\right)/2$$

where $E_{removed}$, $E_{original}$, and $\mu_i$ are the energy of the slab after removing two anions with species $i$, one from each side of the slab, the energy of the slab prior to removing two anions, and the chemical potential of the species $i$ that was removed. The chemical potential was referenced to $H_2$ gas, graphite C, $N_2$ gas, or $O_2$ gas, all at 0 K (no sulfur vacancy formation energies were calculated in this study). A choice of different chemical potential results in an across-the-board constant shift in $E_{vac}$ for the same cation. As an example, when a supercell of the Au nanorod-adsorbed TiC slab contains 126 Ti atoms, 126 C atoms and 36 Au atoms, the C vacancy formation energy is given by

$$\left\{E\left(Ti_{126}C_{124}Au_{36}\right) - E\left(Ti_{126}C_{126}Au_{36}\right) + 2E(C)\right\}/2.$$

The second term is the total energy of the original slab with 1:1 Ti:C ratio plus adsorbed Au, while the first term is that of the slab with 2 C desorbed plus adsorbed Au. The third term is the chemical potential term, which is taken as the formation energy of graphite C in this study. It is possible to, for example, choose the formation energy of diamond C as the reference (chemical potential). In this case, the absolute C vacancy formation energies will change accordingly. However, the relative difference in C vacancy formation energy between, for example, $Ti_{126}C_{126}Au_{36}$ (slab with nanorods) and $Ti_{126}C_{126}$ (slab without nanorods) does not depend at all on the choice of the third chemical potential term.

Defects were separated by at least 9 Å, which exceeds the minimum distance between defects proposed in a previous study [53].

The WF was calculated using the surface-sensitive ionization potential (IP) definition [54], which is simply the difference between the vacuum energy $\varepsilon_{vac}^{surface}$ and the Fermi energy $\varepsilon_{Fermi}^{surface}$ in a slab-and-vacuum model calculation:

$$WF = \varepsilon_{vac}^{surface} - \varepsilon_{Fermi}^{surface}$$

When the compound is a semiconductor or insulator, or in other words, there is an indirect band gap, the IP and electron affinity (EA) were additionally calculated based on the bulk-based definition [54] that excludes the explicit effects of in-gap surface states as in typical IP and EA evaluation[55]. In the bulk-based approach, the IP and EA are obtained by combining surface and bulk calculations as

$$\text{IP} = \varepsilon_{\text{vac}}^{\text{surface}} - \varepsilon_{\text{ref}}^{\text{surface,far}} - \left( \varepsilon_{\text{VBM}}^{\text{bulk}} - \varepsilon_{\text{ref}}^{\text{bulk}} \right)$$

$$\text{EA} = \varepsilon_{\text{vac}}^{\text{surface}} - \varepsilon_{\text{ref}}^{\text{surface,far}} - \left( \varepsilon_{\text{CBM}}^{\text{bulk}} - \varepsilon_{\text{ref}}^{\text{bulk}} \right).$$

A surface calculation using a slab model (supercell) is used to obtain $\varepsilon_{\text{vac}}^{\text{surface}}$ and $\varepsilon_{\text{ref}}^{\text{surface,far}}$, which are the vacuum level and the reference level in the bulk-like region far from the surface, respectively. The bulk-like region is defined as the middle one-third of the slab. On the other hand, a bulk calculation is used to determine $\varepsilon_{\text{VBM}}^{\text{bulk}}$, $\varepsilon_{\text{CBM}}^{\text{bulk}}$, and $\varepsilon_{\text{ref}}^{\text{bulk}}$, which are the VBM, CBM, and the reference level, respectively.

The convergences of the $E_{\text{surf}}$ and WF were checked by building slabs of the same termination with various thicknesses and incrementally increasing the thickness until the changes in the $E_{\text{surf}}$ and WF are less than 5 meV/Å$^2$ and 0.1 eV per inclement, respectively. The resultant slab thickness was larger than 15 Å and three repeat units, and the minimum vacuum thickness is 12 Å. Tables S1-S5 provide information of the slab models, $E_{\text{surf}}$, and WF. The surface orientations were taken with respect to the crystallographic conventional cell as defined in Hinuma *et al.*[56]

## 3. Results and discussion
### 3.1 Work function of investigated compounds
Fig. 3 shows the WF of representative surfaces of considered compounds. Hydrides, carbides, and nitrides are discussed first. The compounds were all metallic except for YN with a very small band gap (0.07 eV) (Tables 1-3). The WFs were all smaller than the WF of Ag (4.56 eV) with the exception of TiC (still close at 4.61 eV), suggesting that many late transition metals, including Cu, Re, and Ru, can act as electron scavengers. As a result, there is a large potential for activation of hydrides, carbides, and nitrides through adsorption of a wide variety of metal nanorod choices.

In contrast, the WF range of oxides and sulfides were large at 1.97 to 8.33 eV and 2.28 to 6.02 eV, respectively. Many oxides and sulfides had a band gap, thus the IP and EA that excludes the explicit effects of in-gap surface states are shown together with the WF in Fig. 4 and 5 for oxides and sulfides, respectively; we note that the band gaps tend to be severely underestimated because the PBEsol functional was used. There are methods to obtain more appropriate band positions that combines the bulk-based approach with hybrid functional and/or $GW$ calculations (details are given in Ref. [55]). In particular, the dielectric-dependent hybrid functional approach, in which the nonlocal Fock exchange mixing is set at the reciprocal of static electronic dielectric constant to improve accuracy [57], is very fast when the hybrid functional calculation is performed non-self-consistently. However, this approach is not applicable to metals because the dielectric constant can be regarded as infinity, thus the Hartree-Fock exchange contribution is zero, resulting in a standard DFT calculation. Regarding WFs, the experimental values of a variety of elementary metals have been reported in the literature. In particular, Derry et al.[58] published recommended values of clean metal surface WFs by orientation. A comparison of the experimental WFs and our calculated WFs are shown in Table S6. The values are very close with a mean error of 0.12 eV. The errors were in the range -0.04 to 0.29 eV for the 10 published WFs. Therefore, the calculated WFs of metals can be considered relatively reliable without further corrections. As a consequence, we did not pursue improvement of band positions to keep the same level of computational approximation over all systems, metallic and non-metallic, at a relatively low cost.

Oxides with a cation valence that is the same as the group number were insulators and have a large WF. In contrast, oxides with reduced cations, in which the formal valence of the cation is smaller than the group number, had a small WF of at most 4.32 eV and were metals or a very narrow band gap semiconductor (case of ScO). This result suggests that choosing a reduced oxide, or at least reducing the surface of an oxide, strongly assists attainment of electron scavenging from nanorod adsorption. The other sulfides other than (Ti, Zr, Hf)$S_2$ and (Ti, Zr, Hf)$S_2$ had WFs of at most 5.07 eV (WF of Re) and most had no or small band gap (band gaps do not exceed 0.10 eV with the exception of $VS_4$ having a relatively large band gap of 0.82 eV). Sulfides with heavily oxidized cations (larger number of S per cation) were low-dimensional and have larger WF than reduced sulfides with three-dimensional structure. In fact, sulfides with larger than 4.30 eV (WF of Zn) were all low-dimensional structures except for $Nb_3S_4$. Therefore, reduced sulfides are likely to be activated by metal nanorod adsorption, which is a trend also found in oxides.

## 3.2 Cation dependence on surface anion vacancy formation energy

Many low-WF systems share the same crystal structure according to Fig. 3. Therefore, the cation dependence on the surface anion vacancy formation energy was evaluated for the fluorite (111) surface in hydrides and the rocksalt (100) surface in nitrides and group 4 and 5 carbides. Some compounds are hypothetical but were included to observe trends over more elements. The correlations with the bulk formation energy $E_{\text{form\_bulk}}$ and WF are shown in Figs. 6(a) and 6(b), respectively. The values in Fig. 6 are listed in Tables. S7-9. $E_{\text{form\_bulk}}$ is defined against the elementary substance of the metal and $H_2$ gas, graphite C, or $N_2$ gas at 0 K. Changing the reference state of H, C, and N results in a constant shift over all $E_{\text{form\_bulk}}$ because the cation to anion ratio is the same. The surface hydrogen vacancy formation energy, $E_{\text{Hvac}}$, of the nine compounds lay in a narrow range between 1.32 and 1.60 eV. The linear fits for hydrides in Fig. 6 appear to be good because of the vertical scale; the coefficient of determination ($R^2$) was actually very bad at 0.14 and 0.03 for $E_{\text{form\_bulk}}$ and WF, respectively. In contrast, $R^2$ of surface carbon and nitrogen vacancy formation energy ($E_{\text{Cvac}}$ and $E_{\text{Nvac}}$, respectively) against $E_{\text{form\_bulk}}$ were very high at 0.96 and 0.81 for carbides and nitrides, respectively. Note that these two $R^2$ are based on different data points (six and nine, respectively), thus these values cannot be directly compared. A relatively high correlation between surface oxygen vacancy formation energy ($E_{\text{Ovac}}$) and $E_{\text{form\_bulk}}$ was also reported in $d^0$ and $d^{10}$ binary oxides in a separate study[53]. $E_{\text{form\_bulk}}$ can therefore be considered as a measure of cation-anion bond strength; atom removal requires severing of bonds, thus correlation of vacancy formation energy and $E_{\text{form\_bulk}}$ has some physical meaning. The band gap also correlated with $E_{\text{Ovac}}$ in $d^0$ and $d^{10}$ binary oxides, but this correlation is meaningless in hydrides, carbides, and nitrides in Fig. 6 because all band gaps are zero. The low correlation between $E_{\text{Hvac}}$, $E_{\text{Cvac}}$, or $E_{\text{Nvac}}$ and WF suggests that, although the WF could be an indicator of the electron scavenger effects, it has only a small contribution to the surface anion vacancy formations. Therefore, explicit calculations containing vacancies are necessary. The correlation for carbides and nitrides was both negative for $E_{\text{bulk}}$ (Fig. 6(a)) but one was positive and the other was negative for WF. One possible reason for the opposite trend in WF is that the WF has very little correlation with $E_{\text{vac}}$, thus the trend can become either positive or negative. In contrast, the correlation between $E_{\text{vac}}$ and $E_{\text{bulk}}$ should be negative, which was the case for all of hydrides, carbides, and nitrides in Fig. 6(a), because a stronger bond, or smaller $E_{\text{form\_bulk}}$, should result in a larger $E_{\text{vac}}$ as more energy is required to break stronger bonds to form a vacancy.

The $E_{Cvac}$ is negative in NbC and VC, and these compounds are experimentally obtained with C deficiencies. Rocksalt $Nb_{1-x}C_x$ forms at roughly 42 to 50 atom% C between room temperature and 2000 °C, and in this range the $Nb_6C_5$ phase is stable below 1050 °C. On the other hand, rocksalt VC, which has the lowest $E_{Cvac}$, is not stable at stoichiometry; the rocksalt phase is stable in a very narrow range at roughly $V_{0.58}C_{0.42}$ at 500 °C. Additional phases are reported, namely $V_6C_5$ and $V_8C_7$, between $V_{0.58}C_{0.42}$ and $V_0C_{1.00}$ below ~1000 °C [40]. Therefore, negative $E_{Cvac}$ in NbC and VC is not a surprising result.

**3.3 Factors to determine surface anion vacancy formation energies and their ML prediction**

Identifying the correlation of the surface anion vacancy formation energies of hydrides, carbides, and nitrides ($E_{(H,C,N)vac}$) with other physicochemical quantities can allow for rationalization and estimation of $E_{(H,C,N)vac}$, which would be beneficial in screening materials for a specific application without significant computational cost. Figure 7 shows the correlations of $E_{(H,C,N)vac}$ with physicochemical properties of hydride, carbide, and nitride compounds appearing in Fig. 6 as well as elemental properties of cation and anion elements for those compounds. We studied correlations across the hydride, carbide, and nitride compounds to find general and versatile relations. For the properties of compounds themselves, $E_{bulk}$ and WF, both of which are obtained by DFT calculations in this study, gave relatively high correlation coefficients although the bulk density and the length of M–X bond, where X is an anion atom to be removed (M–$X_{vac}$ length), showed poor correlations with $E_{(H,C,N)vac}$. This result indicates the importance of not only structural properties but also electronic properties to understand $E_{(H,C,N)vac}$, which is in line with our previous study on surface $E_{Ovac}$ calculations for various metal oxides[53]. Regarding elemental properties, EA and electronegativity for the cations provided relatively high correlation coefficients to the $E_{(H,C,N)vac}$.

Statistical analysis based on machine learning techniques were also carried out to predict $E_{(H,C,N)vac}$ and identify the important factors for their prediction. Descriptors of compounds include $E_{bulk}$ and WF, while elemental descriptors include the EA and electronegativity of the cation and anion. Note that these descriptors were identified thorough testing of several physicochemical properties by considering the prediction accuracy and to reduce multicollinearity. The predictive performance of the ML model based on an extra trees regressor (ETR) [59], a tree ensemble method, was evaluated by Monte Carlo cross validation with 100 times of random leave-25%-out trials. The $R^2$ value obtained was 0.78 and the average root-mean-square error (RMSE) was 0.32 eV, demonstrating relatively high predictive capability of the ML model although the number of datapoints is only 24 in this analysis (Fig. 8a). This result demonstrates that $E_{(H,C,N)vac}$ may be predicted with even higher accuracy without performing computationally-expensive DFT calculations by using readily available descriptors once more data are calculated in the future.

Analysis results obtained using the Shapley additive explanations (SHAP) method[60-62] (version 0.35.0) based on ETR provide further detailed information (Figure 8b). The SHAP approach enables the identification and prioritization of descriptors and thus can be used to explain the contribution of a given input feature to the target ($E_{(H,C,N)vac}$) response. More specifically, the predicted value relative to the dataset average is decomposed into the sum of the contributions of the individual features (descriptors). Rather than considering the importance of a feature to the entire model ('global' explanations), the SHAP values for the features are obtained for individual predictions ('local' explanations). The most important descriptor was identified as $E_{bulk}$, followed by electronegativity for the anion elements as an elemental property, and WF for the $E_{(H,C,N)vac}$ prediction. The analysis also revealed that $E_{(H,C,N)vac}$ (SHAP value) tends to be high when $E_{bulk}$ (feature value) is low.

## 3.4 Reduction in surface vacancy formation energy with metal nanorod adsorption in Ti compounds

Explicit models where metal nanorods with the face-centered cubic (fcc) structure are adsorbed on Ti compounds (support) were prepared and the surface anion formation energies were calculated. Ti was chosen as the cation to be investigated because of its high industrial importance, partly arising from its low price, and existence of relatively high symmetry systems over multiple anion choices. Fig. S1 shows the relation between orientations of a fcc nanorod. Either the {100} or {111} orientation of the nanorod was chosen to be parallel to that of the support surface. A nanorod consisting of six rows in three layers (one, two, and three rows in each layer from the vacuum side) were adsorbed on the support. An additional layer of two rows was adsorbed in $TiH_2$ with the {111} orientation parallel to the support surface. After full relaxation of the support and nanorod coordinates, formation of anion vacancies at the surface resulted in very small displacement of nanorod atoms of typically less than 0.02 Å, although some nanorod atoms displaced by almost 0.06 Å. Therefore, the effect of structural flexibility of the nanorod, which was discussed as an important mechanism to reduce the O vacancy formation energy by Pacchioni and coworkers [28, 31], should be very small in our calculations. Explicit models where metals are adsorbed on other supports could, in principle, be obtained for other systems. However, nanorod adsorption require a relatively small lattice mismatch in the direction parallel to the rod, thereby applicable systems are quite limited. The lattice mismatch problem disappears for very small nanoparticle adsorption on a support, but the unit cell size used in calculation becomes much larger, for instance ~few ten Å along two directions along the slab in a nanoparticle model compared to ~few ten Å and ~ten Å along the direction perpendicular and parallel to the nanorod, respectively, in a nanorod model. In any case, there is a possibility of significant deformation of the nanorod/nanoparticle upon adsorption, as was seen in nanorods some transition metals adsorbed on $In_2O_3$ (111)[43]. Moreover, too strong bonding between atoms of the adsorbate (nanoparticle/nanorod) could result in melting of the nanoparticle/nanorod when adsorbate atoms are more stable when adsorbed to the support compared to when bonding between adsorbate atoms to form a nanoparticle/nanorod.

3.4.1 $TiH_2$

Re and Ru nanorods were able to adsorb with the {100} orientation parallel to the support surface (Fig. 9), while Re, Ru, Rh, and Pd nanorods adsorbed with the {111} orientation parallel to the support surface (Fig. 10). The side view of the model when Ru nanorods were adsorbed are shown in Figs. 9a and 10a, respectively. The support consisted of seven sets of H-Ti-H layers. Activation of H sites near the nanorod, which is represented by the reduction in $E_{Hvac}$ from the value of 1.51 eV without nanorods, were found with all nanorods (Fig. 9b,c, Fig. 10b-d). In Fig. 9, all surface H sites underwent activation, although the extent depended on the site. As expected, H sites near the nanorod were more activated. The WF of Re and Ru slabs are 5.07 and 5.13 eV, respectively, for the most stable (0001) surface in the most stable hcp structure (solid font in Fig. 9). However, these metals are adsorbed as fcc-structure nanorods. The WF as adsorbed nanorods, which were calculated by fixing the atomic coordinates of adsorbed nanorods and then removing the $TiH_2$ support slab, decreased to 4.63 and 4.54 eV for Re and Ru, respectively. Possible causes for the 0.4-0.6 eV difference are the change in structure and surface dependency. The surface dependency on the experimental WF between (10$\bar{1}$0) and (0001) surfaces is as large as 0.80[58]. That being said, the difference between WFs of Re and Ru for both slabs and nanorods differed by less than 0.1 eV, but the $E_{Hvac}$ of the most activated site (0.94 and 0.43, respectively), differed by as much as 0.51 eV. Therefore, the extent of activation could not be precisely determined by the metal WF. In contrast, the closest and next-closest H to each side of the slab was activated when the {111} orientation is parallel to the support surface (Fig. 10). On the other hand, H in the region far from the nanorod was very slightly passivated. The slight differences on passivation with the difference from the nanorod suggests that the effect of local changes in the coordination, especially at the nanorod-support-vacuum boundary, may extend up to over a nanometer or so, but this is not always the case. The WF of the nanorod metal was always larger than the $TiH_2$ support WF for all nanorod metal elements and both WF definitions, meeting the condition for the electron scavenger mechanism.

### 3.4.2 TiC and TiN

Fig. 11 shows Au and Ag nanorods adsorbed on the TiC {100} surface. The slab consisted of seven TiC layers (Fig. 11a). The surface C nearest to the nanorod was separated by three bonds from the nanorod: (Ag or Au) – C immediately below (Ag or Au) – surface Ti – surface C. Passivation of the surface of 0.1-0.2 eV happened (Fig. 11b,c). This may not be surprising for Ag nanorods on TiC because the WF of Ag is marginally smaller than TiC, forcing movement of electrons from the nanorod to the support. However, a similar extent of passivation happened with nanorod adsorption of Au, which has, depending on the definition, 0.7-1.0 eV larger WF than Ag and TiC. Au and Ag nanorods were adsorbed on the TiN {100} surface in a similar manner as on TiC. No activation nor passivation was observed at the surface (Fig. 11d,e).

The absence of activation from the electron scavenger effect in TiN may appear surprising. One possible reason is the distance from the anion site to the nearest metal atom. The distance was 1.83 Å for Re on $TiH_2$ (100), 1.97 Å for Pd on $TiH_2$ (111), and 2.78 Å for Ag on $Ti_2O_3$ (01$\bar{1}$2). In contrast, the distance was 3.94 Å for Ag on TiC (100) and 3.87 Å for Ag on TiN (100). Significant activation happened on O sites with very short distance, typically less than 2.5 Å, from a metal atom of a nanorod adsorbed on $In_2O_3$.[43] The adsorbate metal-anion site distance of close to 4 Å might have been too long to cause activation through the electron scavenger mechanism in TiN.

### 3.4.3 $Ti_2O_3$

Eight types of nanorods, namely Al, Zn, Ag, Re, Ru, Rh, Pd, and Pt in order of increasing WF, were adsorbed on the $Ti_2O_3$ (01$\bar{1}$2) surface (Fig. 12). How the nanorod layers stacked strongly depended on the metal element. Fig. S2 shows the nanorods each viewed from three directions. The O sites were labeled A to F (Fig. 12b). The $E_{Ovac}$ for pair of sites A and F, B and E, and C and D were close to each other.

Fig. 13 compares the $E_{Ovac}$ of sites D, E, and F as a function of the nanorod WF. A clear trend was found where the $E_{Ovac}$ of the D site, which is the closest to a nanorod, decreased with increasing WF, regardless of the WF definition (from a slab in Fig. 13a and nanorods in Fig. 13b). In contrast, the $E_{Ovac}$ of the E site, which is most distant to a nanorod, was very close to the $E_{Ovac}$ without nanorods (5.83 eV) regardless of the nanorod species. The value of $E_{Ovac}$ and the dependence on the nanorod WF for the F site came between sites D and E. The action of the nanorod as an electron scavenger was clearly found and, as expected, the decrease in $E_{Ovac}$ increased with the difference between WFs of the $Ti_2O_3$ support and metal nanorod.

## 3.5 Relation between Bader charge transfer and surface vacancy formation energy with metal nanorod adsorption in Ti compounds

Changes in the Bader charge of the nanorod with surface anion adsorption is a measure of electron scavenger activity by the nanorod. Removal of an anion leaves behind electrons, and part of the charge would be accommodated by the nanorod if the nanorod is an electron scavenger. Charge flow between metal and oxide through WF engineering has been previously proposed, for example by Pacchioni[31], and demonstrated on various metal nanorods adsorbed on the $In_2O_3$ (111) surface[43]. Fig. 14 summarizes the relations between the Bader charge transfer to the nanorod and $E_{vac}$ for various anion vacancy sites in the investigated systems. There is a cluster of points at the intersection of the dotted line, which indicates $E_{vac}$ without nanorods, and zero Bader charge transfer. These points represent anion removal without any interaction with nanorods. The majority of other points lay to the lower left of this cluster, indicating that a decrease in $E_{vac}$ is accompanied by charge transfer to the nanorod (electron scavenger activity by the nanorod). This qualitative trend was regardless of the nanorod element and compound of the support.

## 4 Summary

The WFs of known hydrides, carbides, nitrides, oxides, and sulfides with group 3, 4, or 5 cations were calculated. The WFs of most hydrides, carbides, and nitrides were smaller than the WF of Ag, implying that the electron scavenger effect may occur when late transition metal nanoparticles are adsorbed on the surface. The WF of oxides and sulfides decreased when reduced, suggesting increased activity by reduction. The $E_{vac}$ correlated well with the bulk formation energy in carbides and nitrides, which was also corroborated by statistical analysis using ETR, while almost no correlation was found in hydrides because of the small range of $E_{Hvac}$ values. The electron scavenger effect was explicitly observed in nanorods adsorbed on TiH$_2$ and TiO$_2$; the $E_{vac}$ decreased, accompanied by charge transfer to the nanorod. Activation of hydrides, carbides, and nitrides by nanorod adsorption is expected to open up a new category of supported catalysts.

**Conflicts of interest**
There are no conflicts to declare.

**Acknowledgments**

This study was funded by a grant (No. JPMJCR17J3) from CREST of the Japan Science and Technology Agency (JST), and a Kakenhi Grant-in-Aid (No. 18K04692) from the Japan Society for the Promotion of Science (JSPS). Computing resources of the Research Institute for Information Technology at Kyushu University, ACCMS at Kyoto University, and the Supercomputer Center in the Institute for Solid State Physics at the University of Tokyo were used. The VESTA code [63] was used to draw Figs. 1, 2, 9-12, and S1-S2.

**Supporting Information**
9 supporting tables and 2 supporting figure.

**Tables and Figures**

Table 1. Information on hydrides investigated in this study. MP ID is the Materials Project material ID for the first structure of each prototype. The band gap is the minimum band gap in eV.

| System | Space group | Pearson symbol | Phase | Prototype | MP ID | Band gap | Note |
|---|---|---|---|---|---|---|---|
| $HfH_2$ | $I4/mmm$ | $tI6$ | ε | $ThH_2$ | 27731 | Metal | a) |
| $LaH_3$ | $Fm\bar{3}m$ | $cF16$ | δ | $BiF_3$ | 1018144 | Metal | |
| $NbH_2$ | $Fm\bar{3}m$ | $cF12$ | δ | $CaF_2$ | 24154 | Metal | |
| $ScH_2$ | $Fm\bar{3}m$ | $cF12$ | | $CaF_2$ | | Metal | |
| $TiH_2$ | $I4/mmm$ | $tI6$ | ε | $ThH_2$ | | Metal | b) |
| $V_2H$ | $C2/m$ | $mS6$ | $β_1$ | $AuTe_2$ | 642644 | Metal | |
| $YH_2$ | $Fm\bar{3}m$ | $cF12$ | δ | $CaF_2$ | | Metal | |
| $YH_3$ | $P\bar{3}c1$ | $hP24$ | | $D_3Ho$ | 23706 | Metal | c) |
| $ZrH_2$ | $I4/mmm$ | $tI6$ | ε | $ThH_2$ | | Metal | d) |

a) Reported as Pearson symbol $tI?$ at H composition 64.2% (H-deficient) in the crystal structure data in Ref. [40].
b) According to Ref. [40], the Pearson symbol is $tI2$, and this ε phase is stable below ~40 °C while the δ phase with the $CaF_2$ structure is stable above this critical temperature. The $ThH_2$ structure is a distorted form of the $CaF_2$ structure, thus the ε phase is considered in this study.
c) Reported at composition 74.5 at.% H (H-deficient) in the crystal structure data in Ref. [40].
d) Reported at composition 63.6 at.% H (H-deficient) in the crystal structure data in Ref. [40]. The δ phase with the $CaF_2$ structure is reported as a stable phase between 56.7 to 66.4? at.% (H-deficient) in the crystal structure data in Ref. [40]. The ε phase is considered in this study for the reason in footnote b).

Table 2. Information on carbides investigated in this study. MP ID is the Materials Project material ID for the first structure of each prototype. The band gap is the minimum band gap in eV.

| System | Space group | Pearson symbol | Phase | Prototype | MP ID | Band gap | Note |
|---|---|---|---|---|---|---|---|
| HfC | $Fm\bar{3}m$ | $cF8$ | | NaCl | | Metal | a) |
| La$_2$C$_3$ | $I\bar{4}3d$ | $cI40$ | | Pu$_2$C$_3$ | 1184 | Metal | |
| LaC$_2$ | $I4/mmm$ | $tI6$ | α | CaC$_2$ | 2367 | Metal | |
| Nb$_2$C | $P\bar{3}m1$ | $hP3$ | β | | 2318 | Metal | b) |
| NbC | $Fm\bar{3}m$ | $cF8$ | | NaCl | | Metal | c) |
| Sc$_2$C | $R\bar{3}m$ | $hR3$ | | CdCl$_2$ | n/a | Metal | d) |
| Ta$_2$C | $P\bar{3}m1$ | $hP3$ | α | CdI$_2$ | 7088 | Metal | |
| TaC | $Fm\bar{3}m$ | $cF8$ | | NaCl | | Metal | |
| TiC | $Fm\bar{3}m$ | $cF8$ | | NaCl | | Metal | e) |
| V$_2$C | $Pbcn$ | $oP12$ | α | ζ-Fe$_2$N | 20648 | Metal | |
| Y$_2$C | $R\bar{3}m$ | $hR3$ | | CdCl$_2$ | 1334 | Metal | |
| YC$_2$ | $I4/mmm$ | $tI6$ | α | CaC$_2$ | | Metal | |
| ZrC | $Fm\bar{3}m$ | $cF8$ | | NaCl | | Metal | |

a) Reported at Hf composition 50.5-66% (C-deficient) in the crystal structure data in Ref. [40].
b) Reported as Pearson symbol $hP9$ and space group $P\bar{3}1m$ in the crystal structure data in Ref. [40], but this structure in the Materials Project was used instead.
c) This structure is reported as phase NbC$_{1-x}$ at composition ~41.2 at% C (C-deficient) in the crystal structure data in Ref. [40], but this phase is found at composition 50 at% C in the phase diagram in Ref. [40],
d) This structure is not included in the Materials Project. The sole entry in the Materials Project for Sc$_2$C, as of May 1, 2021, is ID 29941, Pearson symbol $hP3$, space group $P\bar{3}m1$, and has 36 meV/atom higher formation energy than the $hR3$ phase.
e) Reported at C composition 32-48.8% (C-deficient) in the crystal structure data in Ref. [40].

Table 3. Information on nitrides investigated in this study. MP ID is the Materials Project material ID for the first structure of each prototype. The band gap is the minimum band gap in eV.

| System | Space group | Pearson symbol | Phase | Prototype | MP ID | Band gap | Note |
|---|---|---|---|---|---|---|---|
| HfN | $Fm\bar{3}m$ | $cF8$ | | NaCl | | Metal | |
| LaN | $Fm\bar{3}m$ | $cF8$ | | NaCl | | Metal | |
| Nb$_2$N | $P\bar{3}1m$ | $hP9$ | | V$_2$N | 1079585 | Metal | |
| Nb$_4$N$_3$ | $I4/mmm$ | $tI14$ | | N$_3$Nb$_4$ | 569167 | Metal | |
| NbN | $P\bar{6}m2$ | $hP2$ | | WC | 2634 | Metal | a) |
| ScN | $Fm\bar{3}m$ | $cF8$ | | NaCl | | Metal | |
| Ta$_2$N | $P\bar{3}1m$ | $hP9$ | | V$_2$N | | Metal | b) |
| Ti$_2$N | $P4_2/mnm$ | $tP6$ | | Anti-TiO$_2$ (Rutile) | 8282 | Metal | |
| TiN | $Fm\bar{3}m$ | $cF8$ | | NaCl | | Metal | |
| VN | $Fm\bar{3}m$ | $cF8$ | δ | NaCl | | Metal | |
| YN | $Fm\bar{3}m$ | $cF8$ | | NaCl | | 0.07 | |
| ZrN | $Fm\bar{3}m$ | $cF8$ | | NaCl | | Metal | |

a) Two structures are reported in the crystal structure data in Ref. [40]; one is space group $P6_3/mmc$, Pearson symbol $hP8$, prototype AsTi, and MP ID 15799 and the other is space group $P6_3/mmc$, Pearson symbol $hP4$, prototype NiAs, and MP ID 2701. These two structures have formation energies 0.202 and 0.034 eV/atom higher than the MP ID 2634 structure, thus the Materials Project ID 2634 structure is considered in this study.

b) Pearson symbol $hP3$, space group $P6_3/mmc$, prototype Fe$_2$N is also reported in the crystal structure data in Ref. [40]. This $hP3$ structure is not included in the Materials Project as of May 1, 2021.

Table 4. Information on oxides investigated in this study. MP ID is the Materials Project material ID for the first structure of each prototype. The band gap is the minimum band gap in eV.

| System | Space group | Pearson symbol | Phase | Prototype | MP ID | Band gap | Note |
|---|---|---|---|---|---|---|---|
| $HfO_2$ | $P2_1/c$ | $mP12$ | | $ZrO_2$ | 352 | 3.90 | |
| $La_2O_3$ | $Ia\bar{3}$ | $cI80$ | α | $Mn_2O_3$ | 2292 | 3.45 | |
| $La_2O_3$ | $P\bar{3}m1$ | $hP5$ | β | $La_2O_3$ | 1968 | 3.78 | |
| NbO | $Pm\bar{3}m$ | $cP6$ | | NbO | 2311 | Metal | |
| $NbO_2$ | $P4_2/mnm$ | $tP6$ | | $TiO_2$ (Rutile) | 2533 | Metal | b) |
| $Nb_2O_5$ | $C2/c$ | $mS28$ | B(ζ) | a) | 604 | 2.44 | c) |
| ScO | $Fm\bar{3}m$ | $cF8$ | | NaCl | | 2.63 | |
| $Sc_2O_3$ | $Ia\bar{3}$ | $cI80$ | | $Mn_2O_3$ | | 3.75 | |
| $Ta_2O_5$ | $C2/c$ | $mS28$ | B | a) | | 3.07 | d) |
| $Ti_3O$ | $P\bar{3}1c$ | $hP16$ | | | 2591 | Metal | e) |
| $Ti_2O$ | $P\bar{3}m1$ | $hP3$ | | Anti-$CdI_2$ | 1215 | Metal | |
| $Ti_3O_2$ | $R\bar{3}c$ | $hR10$ | | | 978968 | Metal | f) |
| $Ti_2O_3$ | $R\bar{3}c$ | $hR10$ | α,β | α-$Al_2O_3$ | 458 | Metal | g) |
| $TiO_2$ | $I4_1/amd$ | $tI12$ | Anatase | $TiO_2$ (Anatase) | 390 | 1.89 | h) |
| $TiO_2$ | $P4_2/mnm$ | $tP6$ | Rutile | $TiO_2$ (Rutile) | | 1.68 | |
| $V_2O_5$ | $Pmmn$ | $oP14$ | | $V_2O_5$ | 25279 | 1.41 | i) |
| $Y_2O_3$ | $Ia\bar{3}$ | $cI80$ | α | $Mn_2O_3$ | | 4.11 | |
| $ZrO_2$ | $P2_1/c$ | $mP12$ | α | $ZrO_2$ | | 3.58 | |

a) This structure is not found in Ref. [40]. The prototype in the ICSD database [64] is $Sb_2O_3$.
b) Pearson symbol $tI96$, space group $I4_1/a$ is also reported in the crystal structure data in Ref. [40]. This $tI96$ structure was not considered in this study because of its large primitive cell size compared to the $tP6$ phase.

c) This structure was not included in the crystal structure data in Ref. [40]. However, many modifications are reported, including in Ref. [40] ; for details, see ref. Hinuma PRM.

d) "Incommensurate substructures based on an orthorhombic subcell below 1450 °C and on a monoclinic subcell above 1450 °C" in the crystal structure data in Ref. [40]. Many modifications are reported, including in Ref. [40] ; for details, see ref. Hinuma PRM.

e) Reported as Pearson symbol *hP*~16 in the crystal structure data in Ref. [40].

f) Reported as Pearson symbol *hP*~5 and space group *P*6/*mmm* in the crystal structure data in Ref. [40]. The space group *P*6/*mmm* structure is not included in the Materials Project as of May 1, 2021, and the Materials Project ID 978968 structure is considered instead in this study.

g) Both α and β phases have the same Pearson symbol, space group, and prototype in the crystal structure data in Ref. [40]. The Pearson symbol in Ref. [40] is *hR*30, but *hR*10 is the correct Pearson symbol for the α-$Al_2O_3$ structure. The phase diagram in Ref. [40] shows that the β phase is stable between 400 and 1842 °C, and no information is given below 400 °C.

h) Denoted as a metastable phase in Ref. [40].

i) Reported as space group *Pmnm* in the crystal structure data in Ref. [40], which is an unconventional setting of the same space group type as *Pmmn*.

Table 5. Information on sulfides investigated in this study. MP ID is the Materials Project material ID for the first structure of each prototype. The band gap is the minimum band gap in eV.

| System | Space group | Pearson symbol | Phase | Prototype | MP ID | Band gap | Note |
|---|---|---|---|---|---|---|---|
| $Hf_2S$ | $P6_3/mmc$ | $hP6$ | | | 10000 | Metal | a) |
| $HfS_2$ | $P\bar{3}m1$ | $hP3$ | | $CdI_2$ | 985829 | 0.91 | |
| $HfS_3$ | $P2_1/m$ | $mP8$ | | | 9922 | 0.87 | |
| LaS | $Fm\bar{3}m$ | $cF8$ | | NaCl | | 1.26 | |
| $Nb_3S_4$ | $P6_3/m$ | $hP14$ | | $Nb_3Te_4$ | 12627 | Metal | |
| $NbS_3$ | $P2_1/m$ | $mP24$ | α | | 1190583 | Metal | b) |
| ScS | $Fm\bar{3}m$ | $cF8$ | | NaCl | | 0.59 | |
| TiS | $P6_3/mmc$ | $hP4$ | | NiAs | 554462 | Metal | c) |
| $TiS_2$ | $P\bar{3}m1$ | $hP3$ | | $CdI_2$ | | Metal | d) |
| $TiS_3$ | $P2_1/m$ | $mP8$ | | | 9920 | 0.11 | |
| $V_3S$ | $P4_2/nbc$ | $tP32$ | β | | 555283 | Metal | |
| $V_5S_4$ | $I4/m$ | $tI18$ | | $Ta_4Ti_5$ | 1133 | Metal | |
| VS | $Pnma$ | $oP8$ | | MnP | 1868 | Metal | e) |
| $VS_4$ | $C2/c$ | $mI40$ | | | 541155 | 0.83 | f) |
| YS | $Fm\bar{3}m$ | $cF8$ | | NaCl | | 0.96 | |
| $Y_5S_7$ | $C2/m$ | $mC24$ | | $Y_5S_7$ | 15670 | 0.72 | |
| ZrS | $P4/nmm$ | $tP4$ | | γ-CuTi | 7859 | Metal | g) |
| $ZrS_2$ | $P\bar{3}m1$ | $hP3$ | | $CdI_2$ | | 0.77 | |
| $ZrS_3$ | $P2_1/m$ | $mP8$ | | | 9921 | 0.93 | |

a) Reported as space group $P6_3/mmc$ or $P6_3mc$ in the crystal structure data in Ref. [40].
b) Pearson symbol $mP8$, space group $P2_1/m$, and prototype $TiS_3$ as well as Pearson symbol $aP16$, space group $P\bar{1}$, and prototype $NbS_3$ are reported as the crystal structure data for α-$NbS_3$ in Ref. [40]. The former is not in the Materials Project, and the Materials Project ID 978968 structure is considered instead in this study because it has the same space group as the former entry in Ref. [40].
c) Reported at composition 49.7 at.% S (S-deficient) in the crystal structure data in Ref. [40].
d) Reported as space group $P\bar{3}m$ in the crystal structure data in Ref. [40].

e) Reported as space group *Pmma* in the crystal structure data in Ref. [40].
f) Reported as space group *I2/c* in the crystal structure data in Ref. [40] which is an unconventional setting of the same space group type as *C2/c*.
g) Pearson symbol *tP*4, space group *P4/nnm*, and prototype γ-CuTi is also reported as the crystal structure data for ZrS in Ref. [40].

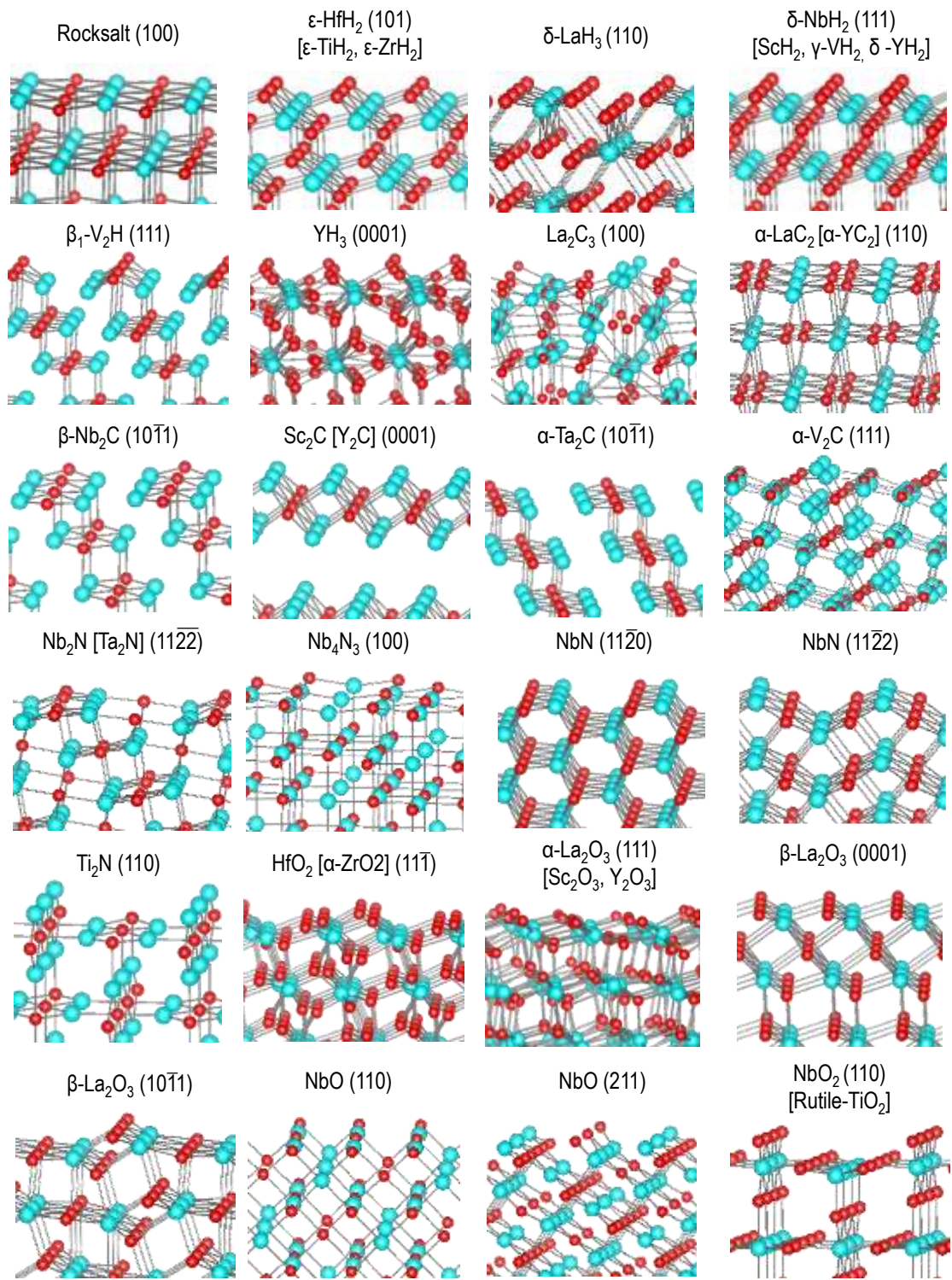

Fig. 1. Representative surfaces of compounds investigated in this study. Blue and red circles indicate cations and anions, respectively.

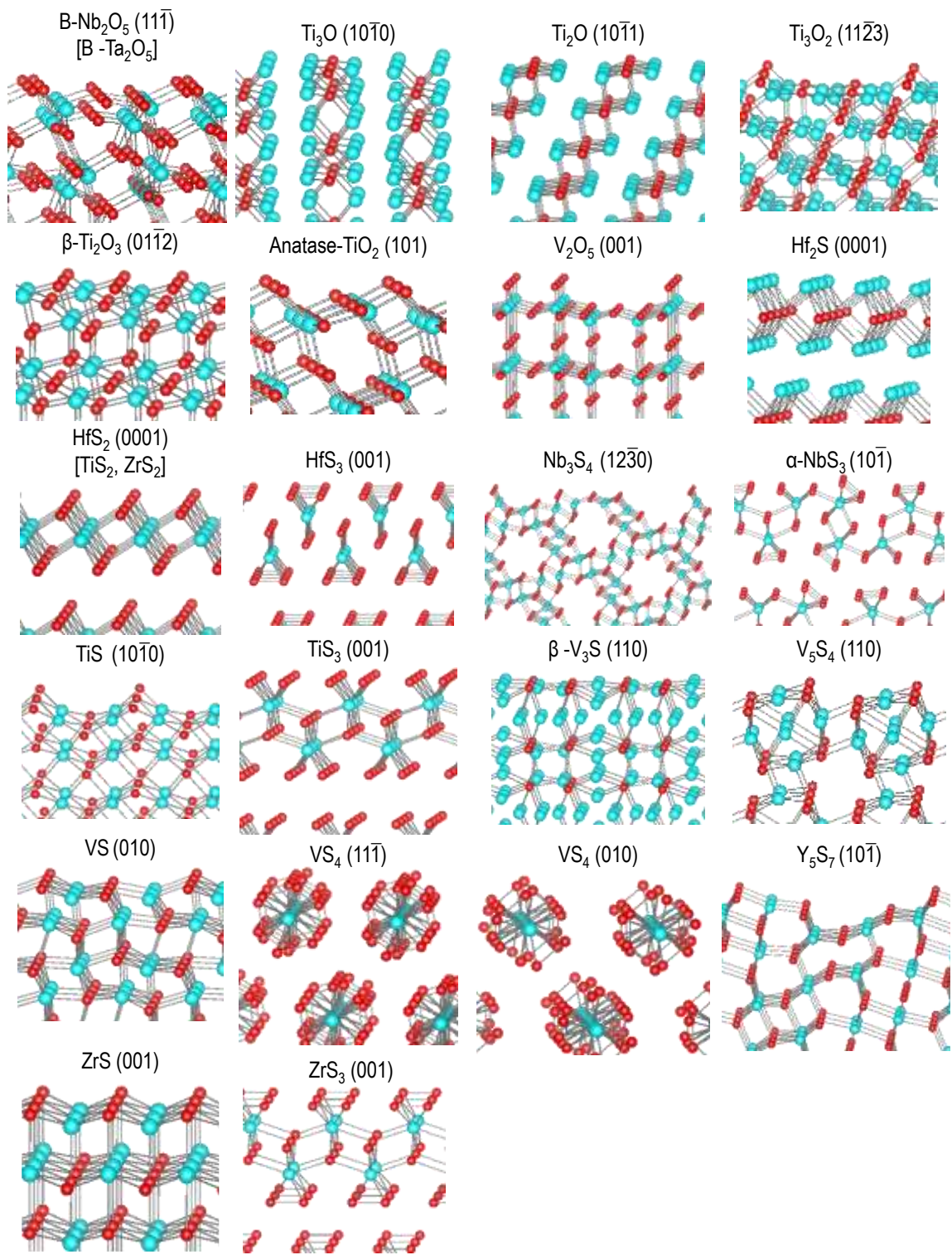

Fig. 2. Representative surfaces of compounds investigated in this study. Blue and red circles indicate cations and anions, respectively.

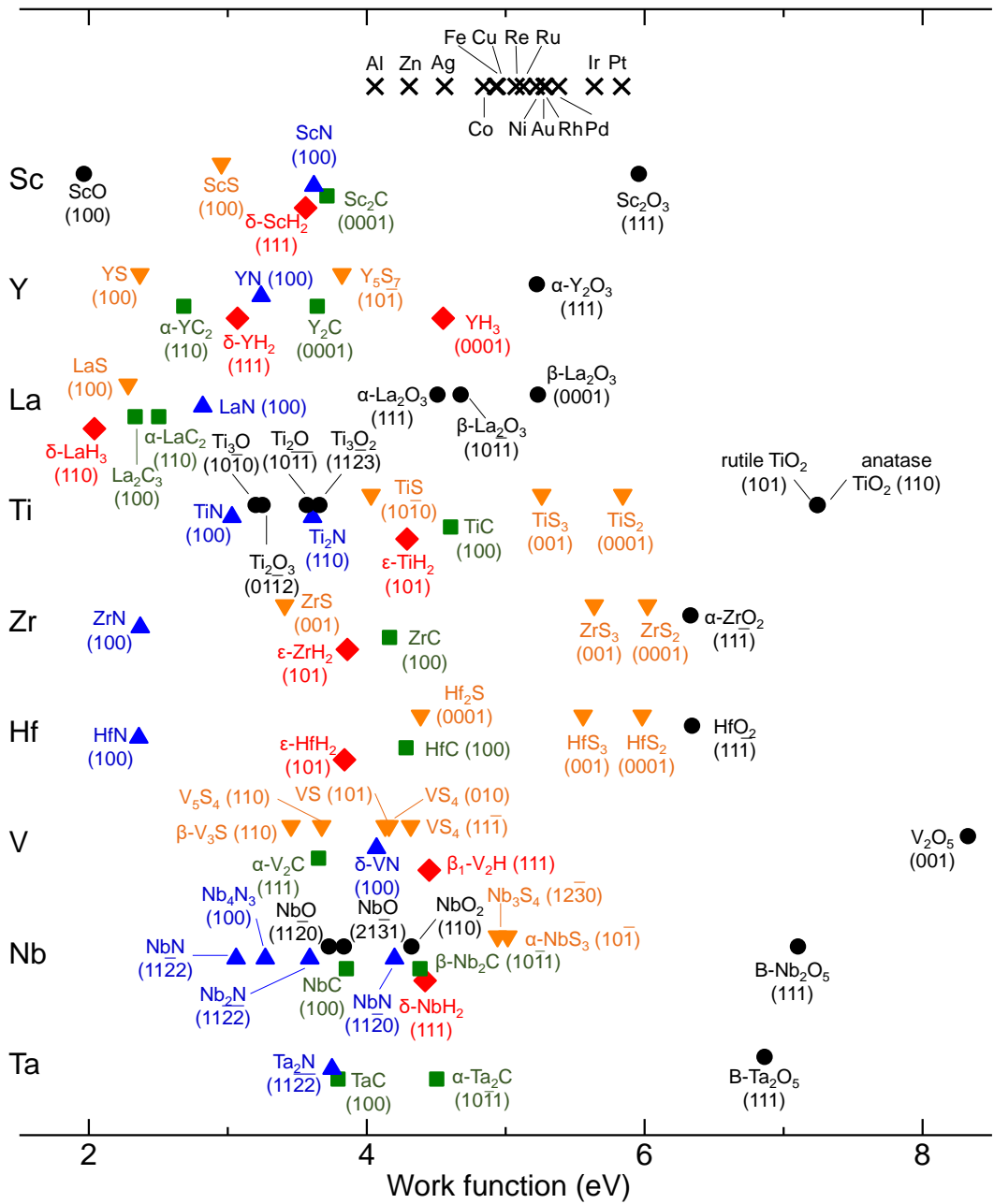

Fig. 3. WF of the representative surface of various metal elements and compounds.

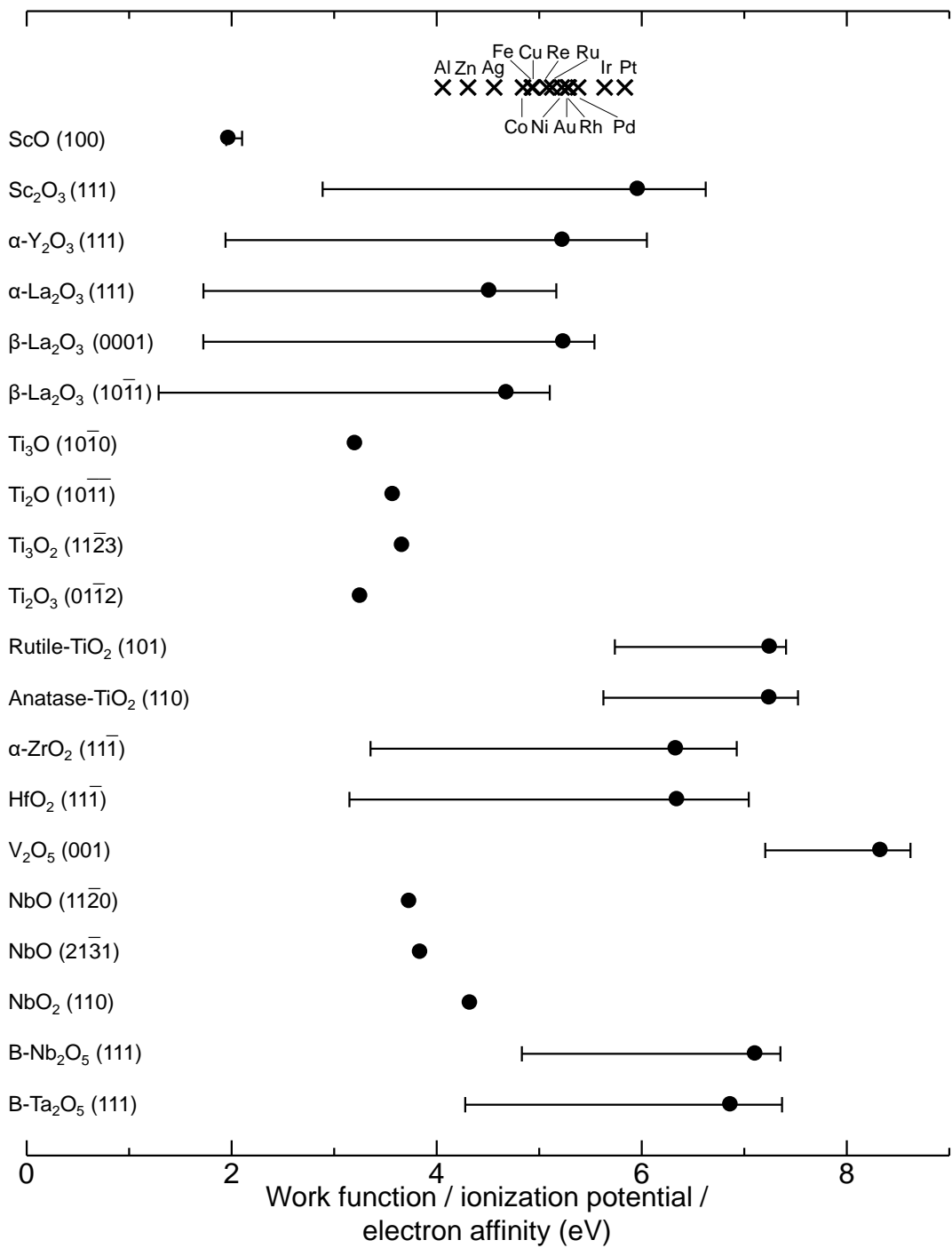

Fig. 4. WF (circles, defined as the surface-sensitive IP [54]) as well as the IP and EA based on the bulk-based definition [54] ( ⊢•⊣ ) of the representative surface of various oxides. The WF of metal elements (crosses) are also shown.

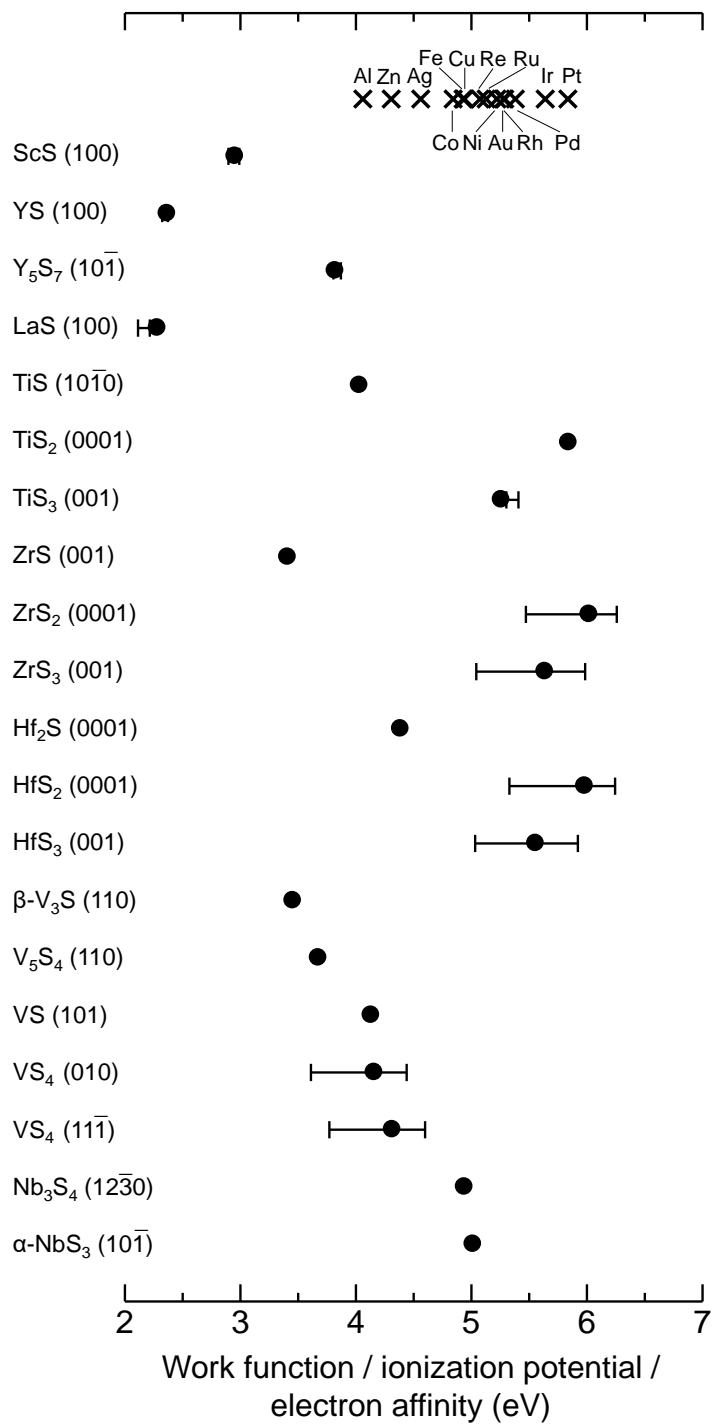

Fig. 5. WF (circles, defined as the surface-sensitive IP [54]) as well as the IP and EA based on the bulk-based definition[54] (bar) of the representative surface of various sulfides. The WF of metal elements (crosses) are also shown.

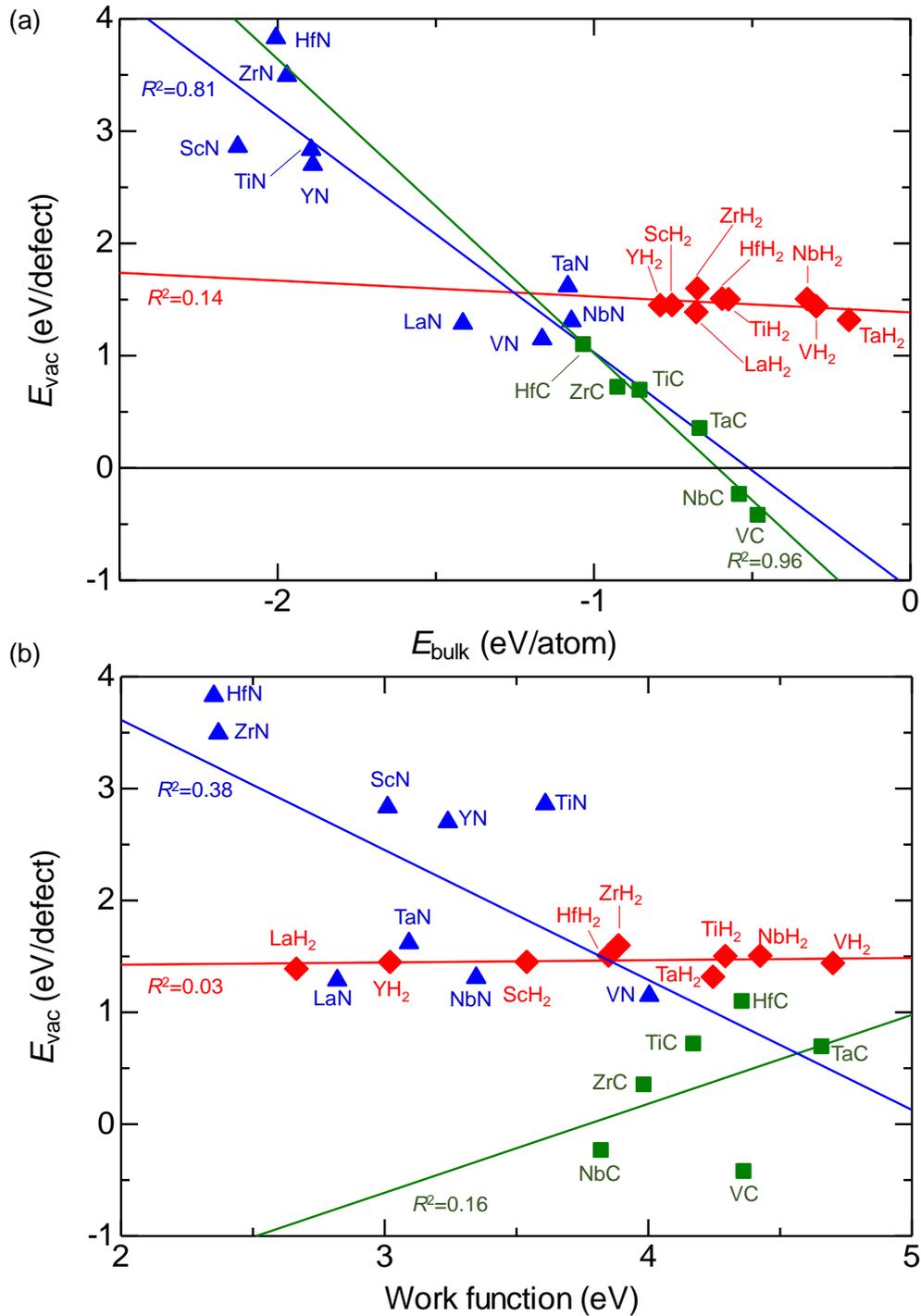

Fig. 6. Surface anion vacancy formation energy ($E_{vac}$) versus (a) bulk formation energy ($E_{bulk}$) and (b) WF. Hydrides have the $CaF_2$ (fluorite) or $ThH_2$ (distorted fluorite) structure, and the $CaF_2$ structure (111) surface or the equivalent (101) surface in the $ThH_2$ structure is considered. Carbides and nitrides have the NaCl (rocksalt) structure, and the (100) surface is considered. Some compounds are hypothetical.

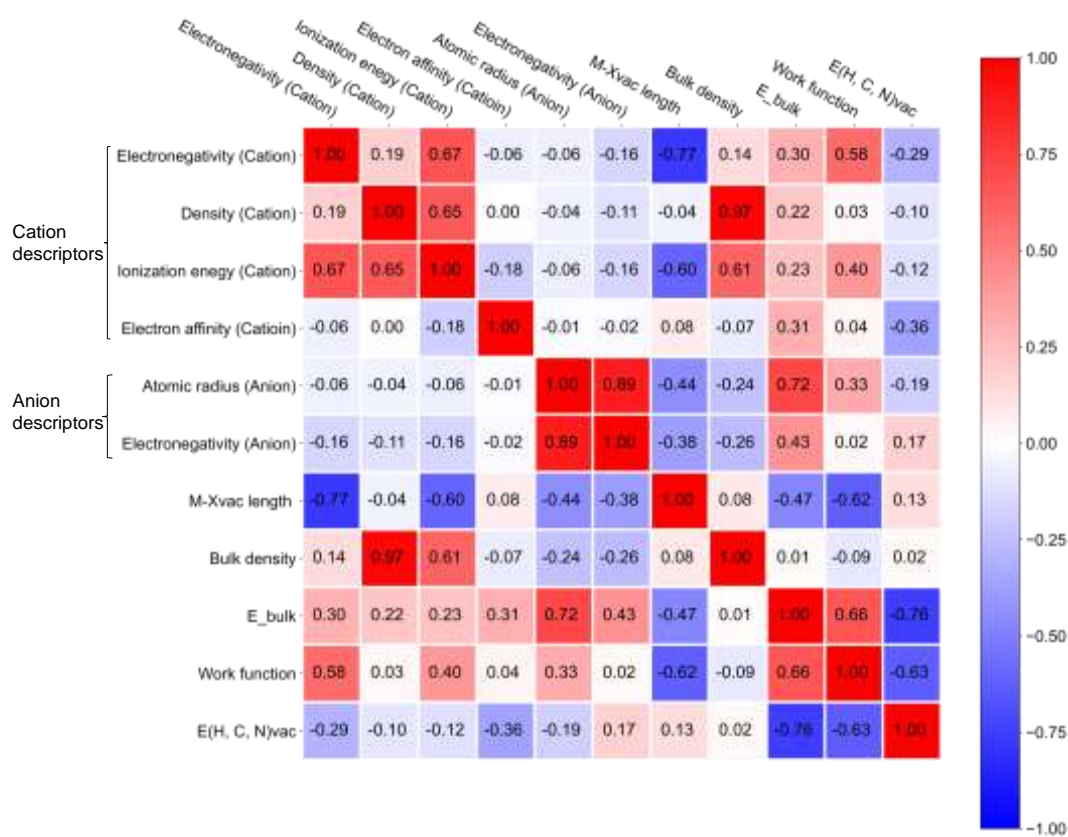

Fig. 7. Correlation map for $E_{(H,C,N)vac}$, physicochemical properties of hydride, carbide, and nitride compounds (MX) in Fig. 6, and elemental properties of cation and anion elements for those compounds (M and X, respectively). The physicochemical properties of hydride, carbide, and nitride compounds include bulk density, $E_{bulk}$, WF, and length of M–X bond where X is an anion atom which will be removed. The latter three quantities were obtained by DFT calculations in this study. The correlation coefficients ($R$) are indicated by the numbers in the squares.

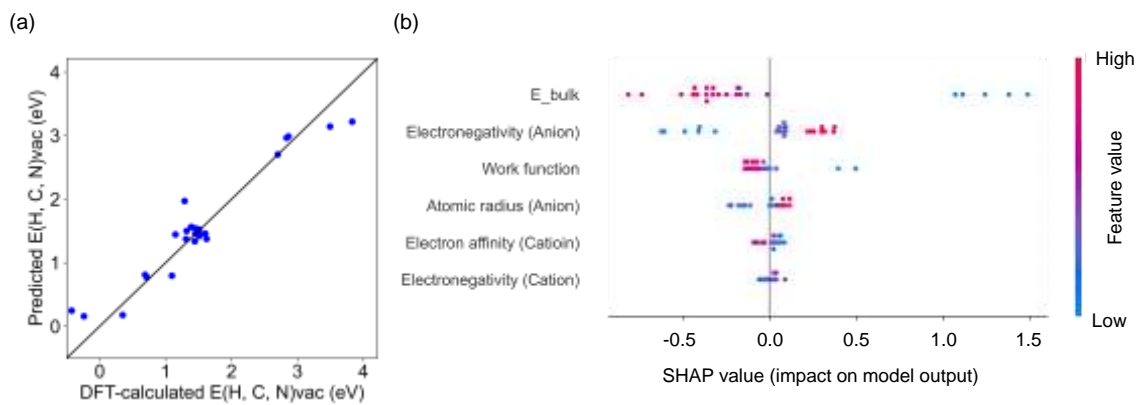

Fig. 8. (a) The out-of-sample prediction performance by 100 times of random leave-25%-out cross validation; DFT-calculated $E_{(H,C,N)vac}$ and values predicted using ETR. (b) SHAP values of the six descriptors in predicting $E_{(H,C,N)vac}$ using ETR. SHAP values for individual factors are plotted as dots (blue corresponds to low features, red to high features). Here, features are ordered in descending order according to the sum of the absolute values of the SHAP values (importance of the descriptors for prediction). Dots are displaced vertically to reflect the density of datapoints at a given SHAP value.

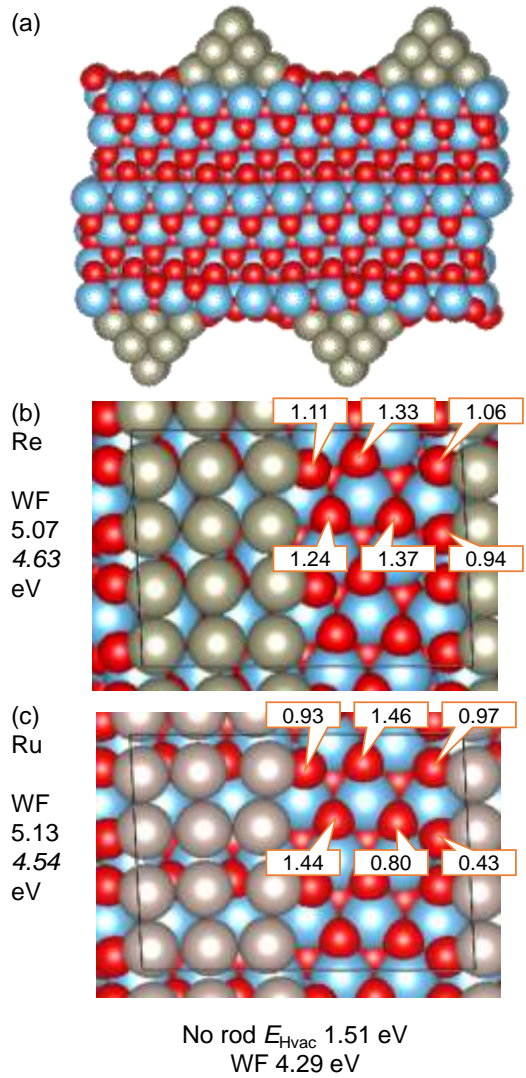

Fig. 9. Adsorption of a nanorod of a face-centered cubic metal on the TiH$_2$ (101) surface (corresponds to the fluorite (111) surface). The {100} orientation of the nanorod is parallel to the slab surface. (a) Side view. Bottom row of the (b) Re and (c) Ru nanorod shown together with the slab surface. The numbers are $E_{Hvac}$ of the corresponding H atom in eV/defect. Circles in blue, red, and the other color indicate Ti, H, and Pd or Ru, respectively. The WF in solid and italic letters are those of the slab and nanorod, respectively. The atomic coordinates of the nanorod used to obtain the WF are those of the nanorod on the slab without anion removal.

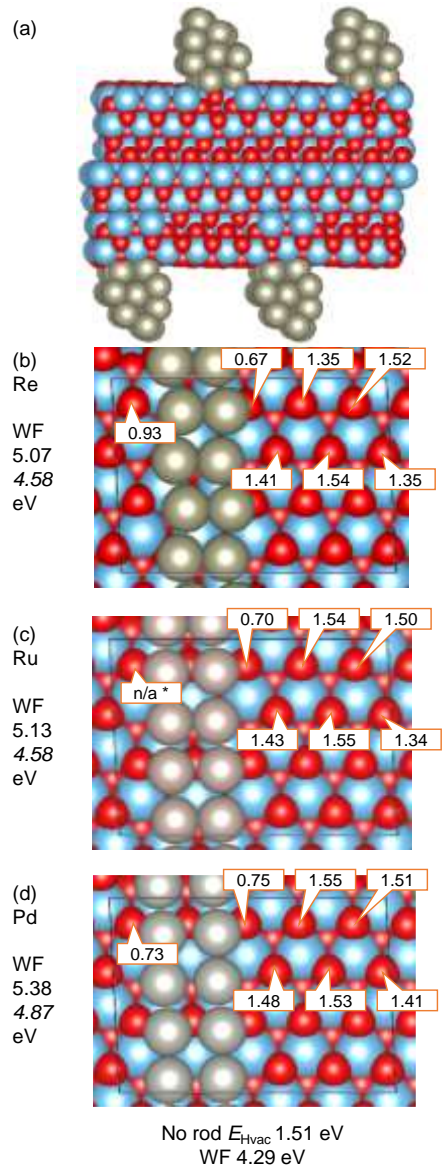

Fig. 10. Adsorption of a nanorod of a face-centered cubic metal on the TiH$_2$ (101) surface (corresponds to the fluorite (111) surface). The {111} orientation of the nanorod is parallel to the slab surface. (a) Side view. Bottom row of the (b) Re, (c) Ru, or (d) Pd nanorod shown together with the slab surface. The numbers are $E_{Hvac}$ of the corresponding H atom in eV/defect. Circles in blue, red, and the other color indicate Ti, H, and the metal nanorod element, respectively. The box indicates the supercell used for $E_{vac}$ calculation. Note for *: excessive deformation of some atoms was observed, thus no $E_{Hvac}$ is provided. The WF in solid and italic letters are those of the slab and nanorod, respectively. The atomic coordinates of the nanorod used to obtain the WF are those of the nanorod on the slab without anion removal.

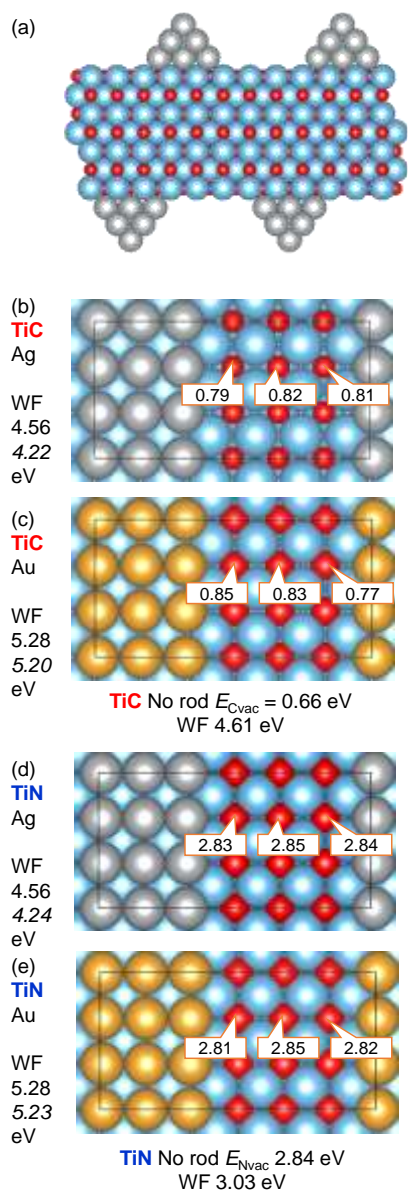

Fig. 11. Adsorption of a nanorod of a face-centered cubic metal on the rocksalt (100) surface. The {100 orientation of the nanorod is parallel to the slab surface. (a) Side view. Bottom row of the (b,d) Ag or (c,e) Au nanorod shown together with the (b,c) TiC or (d,e) TiN slab surface. The numbers are (b,c) $E_{Cvac}$ or (d,e) $E_{Nvac}$ of the corresponding C or N atom in eV/defect. Circles in blue, red, and the other color indicate Ti, C or N, and Ag or Au, respectively. The box indicates the supercell used for $E_{vac}$ calculation. The WF in solid and italic letters are those of the slab and nanorod, respectively. The atomic coordinates of the nanorod used to obtain the WF are those of the nanorod on the slab without anion removal.

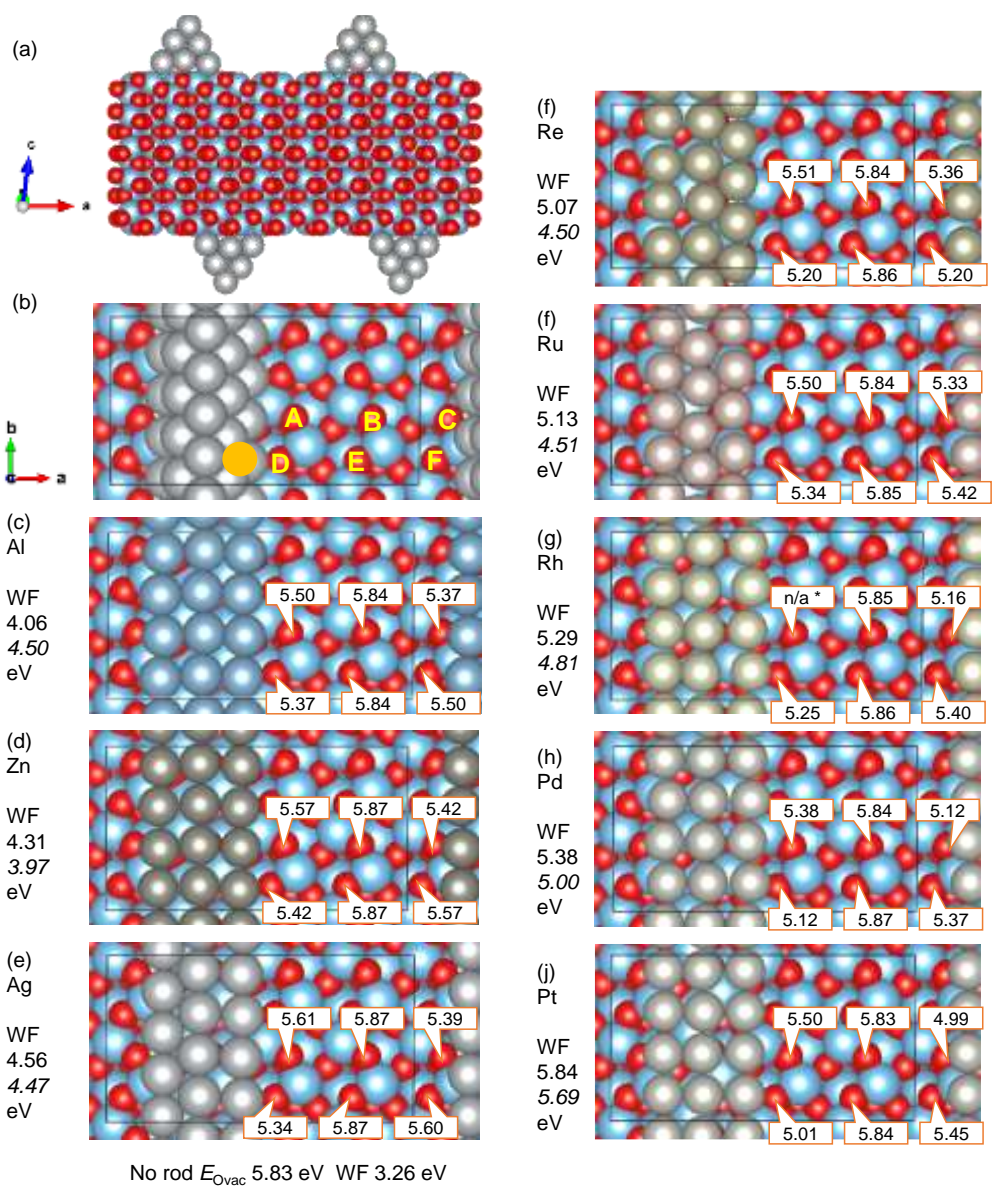

Fig. 12. Adsorption of a nanorod of a face-centered cubic metal on the $Ti_2O_3$ $(01\bar{1}2)$ surface. (a) Side view. (b) Top view and labels of O sites. The numbers in (c-j) are $E_{Ovac}$ of the corresponding O atom in eV/defect. Circles in blue, red, and the other color indicate Ti, O, and the metal nanorod element, respectively. The box indicates the supercell used for $E_{Ovac}$ calculation. Note for *: excessive deformation of some atoms was observed, thus no $E_{Ovac}$ is provided. The WF in solid and italic letters are those of the slab and nanorod, respectively. The atomic coordinates of the nanorod used to obtain the WF are those of the nanorod on the slab without anion removal.

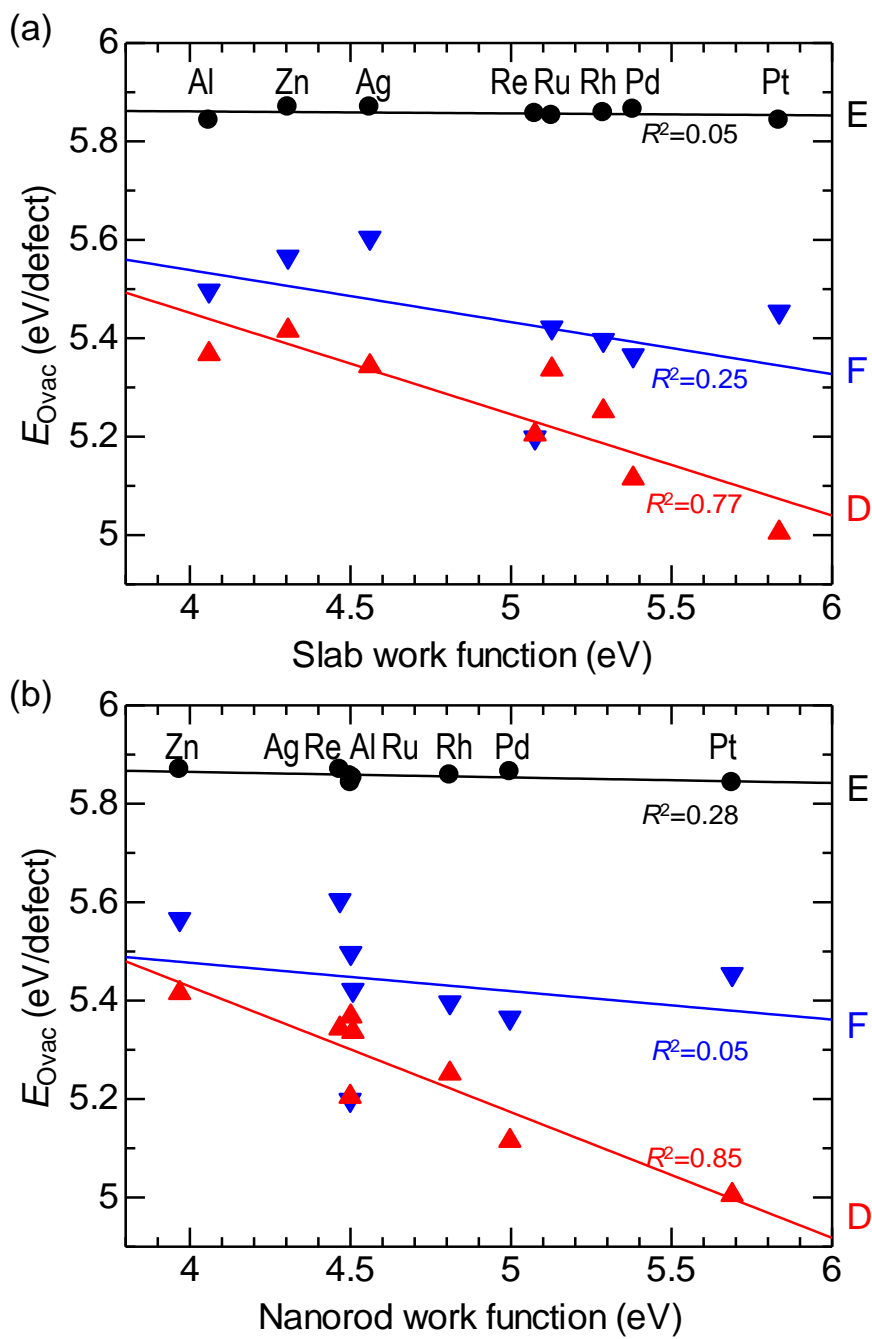

Fig. 13. Relation between $E_{Ovac}$ for O sites D, E, and F and WF in Fig. 12. The WF is from (a) a slab of the most stable surface in the most stable structure in Table S6 and (b) the nanorod with the atomic configuration in Fig. S2 but without the $Ti_2O_3$ support.

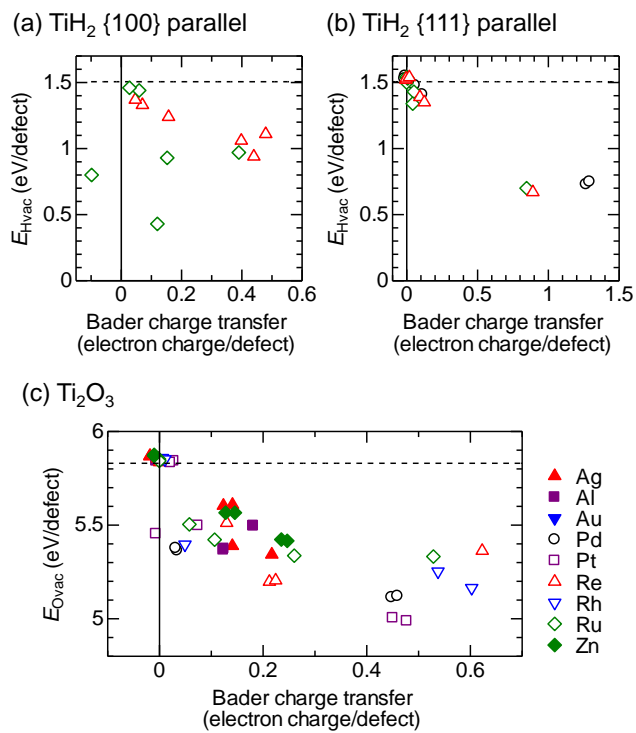

Fig. 14. Relation between $E_{vac}$ and Bader charge transfer to the nanorod upon surface anion removal. A positive value in the horizontal axis means that the nanorod gains electrons (reduced) with anion removal. The dotted line indicates $E_{vac}$ without nanorods.

TOC graphic

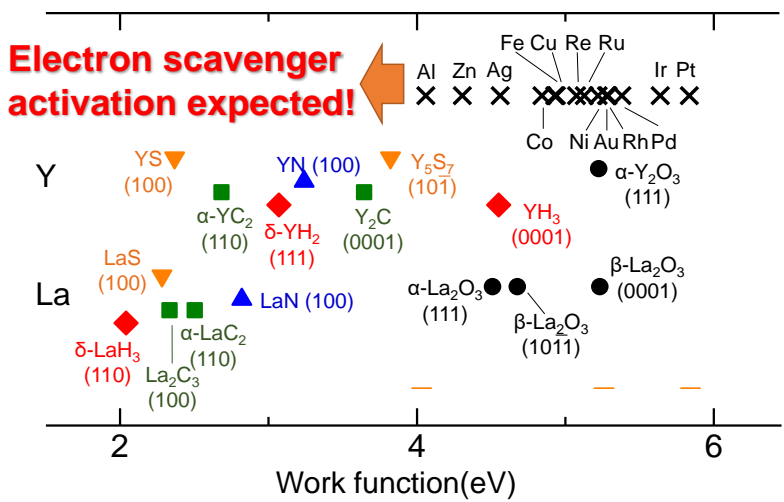

Supports with low work function have the potential to be activated by metal nanoparticle adsorption via the electron scavenger effect.